\documentclass[sigconf]{acmart}
\AtBeginDocument{%
  }

\setcopyright{acmlicensed}
\copyrightyear{2018}
\acmYear{2018}
\acmDOI{XXXXXXX.XXXXXXX}
\acmConference[Conference acronym 'XX]{Make sure to enter the correct
  conference title from your rights confirmation email}{June 03--05,
  2018}{Woodstock, NY}
\acmISBN{978-1-4503-XXXX-X/2018/06}

\acmSubmissionID{3066}

\usepackage{multirow}
\usepackage{array}
\usepackage{colortbl}
\usepackage{ragged2e}

\begin{document}

\title{Working Through Representation: Malleable Workspaces for Human--Agent Knowledge Work}

\author{Nan Chen}
\affiliation{%
  \institution{Microsoft Research Asia}
  \city{Shanghai}
  \country{China}}
\email{nanchen@microsoft.com}

\author{Long Ling}
\affiliation{%
  \institution{Tongji University}
  \city{Shanghai}
  \country{China}}
\email{lucyling0224@gmail.com}

\author{Luna K. Qiu}
\affiliation{%
  \institution{Microsoft Research Asia}
  \city{Shanghai}
  \country{China}}
\email{lunaqiu@microsoft.com}

\author{Zilong Wang}
\affiliation{%
  \institution{CataX AI}
  \city{Shanghai}
  \country{China}}
\email{wangzilong@cata-x.ai}

\author{Yuqing Yang}
\affiliation{%
  \institution{Microsoft Research Asia}
  \city{Shanghai}
  \country{China}}
\email{yuqyang@microsoft.com}
\renewcommand{\shortauthors}{Trovato et al.}

\begin{abstract}
  Generative AI supports knowledge work, yet conversational histories may not capture the evolving organization that guides subsequent action. We introduce the \textbf{Malleable Workspace}, a conceptual model of how humans and agents form and revise a Work Representation and use its organization to advance work within a wider Work Environment. We developed it iteratively through conceptual analysis, comparison of 58 papers, and the construction and three-month longitudinal study of MalleableSpace with seven experienced agent users. Systems varied in what they represented as units and relations. System-assembled context was common; more systems supported local incorporation of agent outputs than reorganization, and fewer supported consequential agent-driven changes beyond the representation. The field study showed how representational roles and organization developed in use. These observations refined the model's account of how representation-grounded action depends on and can change the Work Environment. We contribute a conceptual and analytic foundation for representation-centered human--agent design.
\end{abstract}



\begin{teaserfigure}
  \centering
  \includegraphics[width=0.8\textwidth]{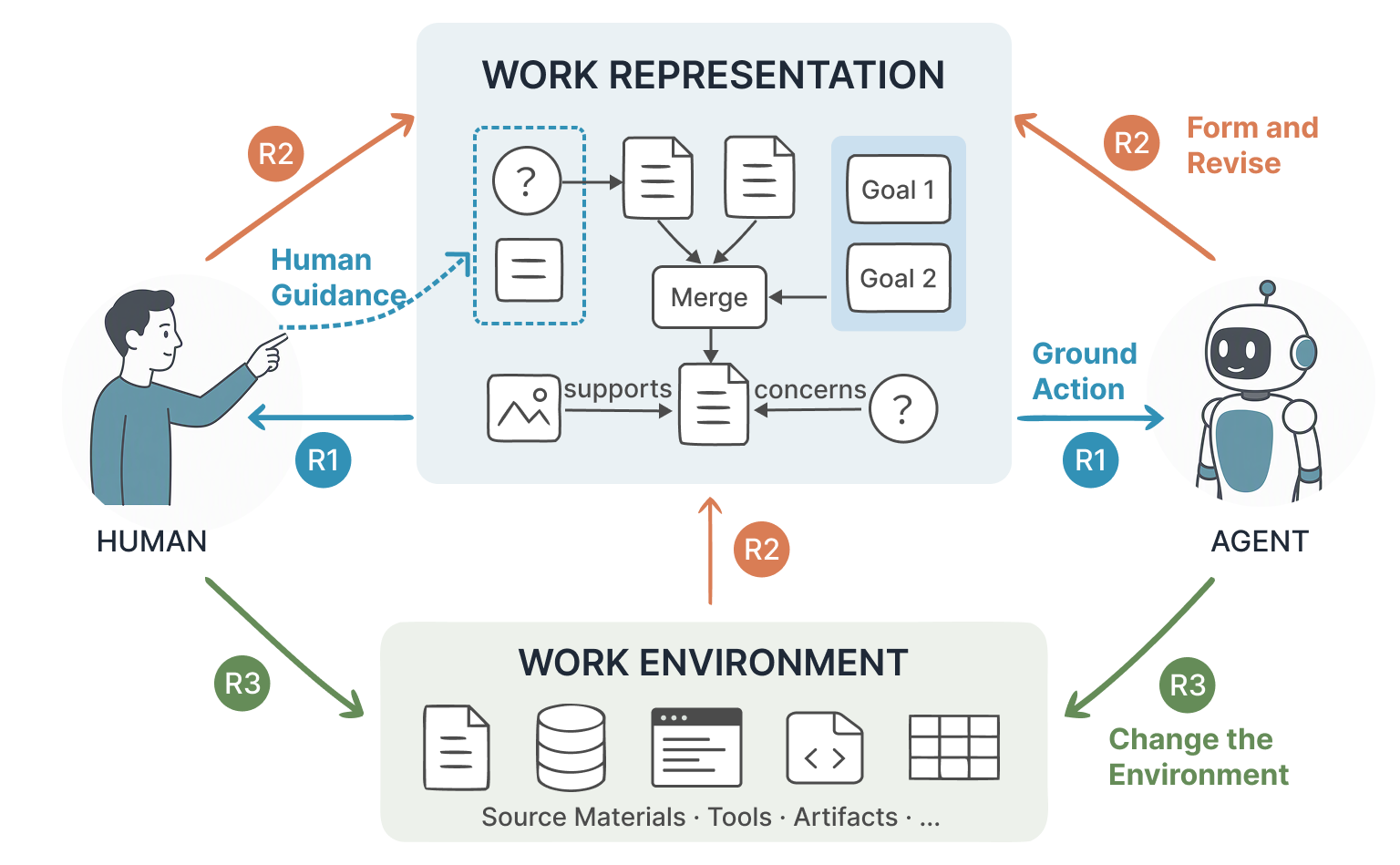}
  \caption{\textbf{The Malleable Workspace: working through an evolving representation.} The Work Representation organizes units---such as documents, questions, goals, and operations---and relations among them. Its organization grounds human and agent action (R1), with human guidance directing particular agent actions. Humans and agents selectively interpret relevant materials and outcomes, forming and revising the representation (R2). Representation-grounded action can produce consequential changes beyond the representation in the Work Environment (R3).}
  \Description{A human and an agent flank a Work Representation containing document, image, question, goal, and merge-operation units. Labelled links, grouped goals, and input/output paths illustrate relations. R1 connects the representation to human and agent action; human guidance points to selected units. R2 connects the human, agent, and Work Environment to selective interpretation and the formation and revision of units and relations. R3 connects human and agent action to changes in the Work Environment below, illustrated by source materials, tools, and artifacts.}
  \label{fig:malleable-workspace-model}
\end{teaserfigure}

\received{20 February 2007}
\received[revised]{12 March 2009}
\received[accepted]{5 June 2009}

\maketitle

\section{Introduction}
\label{sec-intro}
Generative AI is increasingly used in knowledge work to support people in gathering and synthesizing information, exploring interpretations and alternatives, and producing or revising artifacts~\cite{Suh2023_5M75CFML,Fok2024_9G6DRD7Z,Kang2025_FK9XX3VC,Zhang2023_A9FG546C}. Within this work, people decide which materials and AI-generated outputs matter and establish relations among them, such as grouping ideas by theme or linking claims to supporting evidence~\cite{kidd1994marks,Russell1993_MLTWGDTN,Pirolli2005_X2GDEDPH}. External representations, such as notes grouped on a canvas or outlines organizing a draft, make these materials and their relations available for inspection and revision~\cite{larkin1987diagram,norman1991cognitive,zhang1994representations,scaife1996external}. As work progresses, people revise these representations to express their evolving understanding and use them to develop new insights and guide subsequent action~\cite{klein2006sensemaking2,schon1983reflective,dorst2001creativity}.

Such AI-supported knowledge work commonly unfolds through conversational interfaces, which organize human--AI interaction as sequences of turns~\cite{sun2024generativewild,peng2025layusers}. Successive exchanges support the articulation and refinement of ideas, while branching allows users to develop alternative conversational paths~\cite{suchmann2023branching,zhang2025mindalogue,li2026branchat}. Conversational histories preserve how interaction unfolded, but they do not necessarily show how relevant materials and AI-generated outputs currently relate or what these relations imply for subsequent action. Users may therefore need to reconstruct and update this understanding as work progresses~\cite{zhang2018groupsensemaking,sabei2024professionalim,smith2000conversationtrees,zhang2025mindalogue}.

Recent human--AI systems move beyond chat by organizing work in editable visual representations, such as canvases, node-link diagrams, hierarchical outlines, and tables~\cite{Jiang2023_IGH9MP27,Suh2023_5M75CFML,Kang2023_Synergi,Chen2025_MeetMap,Chen2026_ThinkFlow,Zhou2026_JCPKGCB6}. These representations can persist as human--AI work continues, be reorganized by humans and agents, and ground subsequent model action~\cite{Zhang2023_A9FG546C,Tang2026_LECYXG6A,Li2026_YFH7U76P}.
% Previous closing sentence of the third paragraph:
% To advance their work, humans and agents need to revise this organization as their understanding develops and use it to guide subsequent action.
These developments raise a conceptual question: how can an evolving representation mediate ongoing collaboration between humans and agents?
% Understanding how representations mediate collaboration in these systems can inform the design of future human-agent interactions. We therefore ask: how can an evolving representation mediate ongoing collaboration between humans and agents?

% Previous fourth paragraph:
% Existing models explain aspects of this representation-mediated activity. Models of instruction formation and AI instruments describe how represented elements express intent and shape generative action \cite{Shen2026_IAI,Riche2025_AIInstruments}. Other accounts foreground adaptable collaborative processes, linked analytical artifacts, and shared evidence and interpretations \cite{Wang2025_IPI,Wang2026_FiguresAsInterfaces,Guan2026_SharedRepresentationalWorkspace}. However, these models do not fully explain how an evolving representation is formed and revised by humans and agents over time, how its organization grounds subsequent action, and how that action becomes consequential in the wider work environment.
Existing accounts of human--AI work offer important foundations for addressing this question. Models of instruction formation and AI instruments describe how represented elements express intent and shape generative action~\cite{Shen2026_IAI,Riche2025_AIInstruments}. Other accounts foreground adaptable collaborative processes, analytical artifacts as interfaces for further action, and shared workspaces for organizing evidence and interpretations~\cite{Wang2025_IPI,Wang2026_FiguresAsInterfaces,Guan2026_SharedRepresentationalWorkspace}.

% Previous opening sentence of the model introduction:
% We address this gap with the \textbf{Malleable Workspace} (Figure~\ref{fig:malleable-workspace-model}), a conceptual model centered on an evolving \textbf{Work Representation} that humans and agents can revise as work develops and use to guide subsequent action.
Building on these foundations, we introduce the \textbf{Malleable Workspace} (Figure~\ref{fig:malleable-workspace-model}),  a conceptual model that explains how an evolving \textbf{Work Representation} mediates ongoing human-agent collaboration and how this collaboration both draws on and changes the broader work environment. The representation organizes \textbf{units}, such as claims, evidence items, or drafts, and \textbf{relations} among them to express a current, revisable state of work. Humans and agents can reshape this representation as work develops. Its \emph{malleability} allows humans to express and revise their understanding of the work. The resulting organization, together with human guidance, directs agent action, reducing the need to reconstruct relevant context in successive prompts. Such representation-grounded action can change the broader \textbf{Work Environment}, which includes the tools, source materials, artifacts, and other resources involved in the work. Humans and agents also interpret relevant materials and outcomes from this environment and selectively incorporate them into the representation, updating it to guide further action.

We developed the Malleable Workspace iteratively through conceptual analysis, cross-system comparison, and the construction and study of \textbf{MalleableSpace}, a spatial human--agent system operationalizing the emerging model. Four structural mismatches---unit, relation, state, and progression---motivated an initial model that informed both our analysis of 58 recent papers on human--agent systems and the design of MalleableSpace. We then conducted an exploratory three-month longitudinal field study with seven experienced agent users who incorporated MalleableSpace into self-selected work. Insights from the comparison and study prompted model revisions and renewed corpus analysis, sharpening its account of the Work Environment and how representation-grounded action can change that environment.

Across the corpus, systems differed in what they made addressable as units and in the interpretive, compositional, and developmental continuity preserved through relations. Systems commonly supported system-assembled context, while agent-directed acquisition was less common. Local incorporation of agent outputs was widely supported, whereas reorganization of existing units was less prevalent. Most systems supported agent-driven changes within the Work Representation; a smaller subset used represented organization to guide consequential changes to distinct artifacts or environments beyond it. Together, these patterns showed that the model can describe and distinguish representational configurations across diverse interfaces and identify the boundaries of their support for continued human--agent work.

The longitudinal study of MalleableSpace elaborated how a Work Representation functioned in continued use. The same representational forms supported multiple roles as participants put them to use in different ways, while represented organization was progressively formed and revised. Using this represented organization in work beyond MalleableSpace initially required participants to carry context into their existing tools and agents. Connecting the Space to these agents enabled represented organization to guide changes to external artifacts. Relevant materials and outcomes could be selectively incorporated into the representation through subsequent work. These observations motivated the inclusion of the Work Environment as a core model component and sharpened the model's account of how representation-grounded action both depends on and can change that environment.

Drawing on the model and both investigations, we identify four design considerations for supporting continued work through revisable representations. Taken together, this work makes three contributions:

\begin{itemize}

\item We introduce the \textbf{Malleable Workspace}, a conceptual model centered on an evolving Work Representation through which humans and agents organize and revise relevant units and relations and ground subsequent action within a broader Work Environment.

\item We apply the model to analyze 58 recent papers on human--agent systems, showing its utility for describing and distinguishing how systems organize evolving Work Representations and connect them to continued action, while revealing recurring patterns and underexplored design possibilities.

\item We operationalize and interrogate the model by constructing \textbf{MalleableSpace}\footnote{Source code:
\url{https://github.com/microsoft/huabu}.} and conducting a longitudinal field study of its use. The study shows how representational roles and organization develop in continued work and refines the model's account of how the Work Representation draws on and guides changes to the broader Work Environment.

\end{itemize}

\section{Related Work}
\label{sec-related}
Our work builds on three strands of prior research: conversation history and continuity, external representations in knowledge work, and interaction models for human--AI work.

\subsection{Conversation History and Continuity of Work}
\label{sec:conversational-continuity}

AI-supported knowledge work often involves repeated exchanges with LLMs as users refine their intentions. Studies of creative practice and information seeking document how users assess outputs, clarify needs, and explore alternatives~\cite{sun2024generativewild,peng2025layusers}. Conventional chat organizes these exchanges as a linear sequence of turns, each comprising one user or LLM contribution. Conversational branching lets users explore alternative paths from an earlier turn, reusing the preceding history while preserving the original trajectory~\cite{suchmann2023branching,li2026branchat,anthropic2026claudesessions,microsoft2026vscodeforking}.

Work-relevant information can become distributed across turns and paths, within a single chat or across separate chats and branches. Research on human--human messaging describes the effort of recovering prior context and connecting relevant but nonadjacent messages~\cite{zhang2018groupsensemaking,sabei2024professionalim,smith2000conversationtrees}. Related challenges also arise in conversations with LLMs, including navigating long conversations, locating variations, and comparing outputs~\cite{zhang2025mindalogue,reza2024abscribe,gero2024sensemaking}. Prior work supports continuity through reuse of past conversations and retained memories~\cite{cox2023previoussessions,jo2024longtermmemory}, with user-facing tools for inspecting, editing, or explicitly reusing that information~\cite{huang2023memorysandbox,yen2024memolet}. The relevance of retained information can change as users' intentions evolve~\cite{jones2025agentmemory,vaithilingam2025semanticcommit}. This work motivates examining how information distributed across conversational histories can be brought together and reorganized to support changing work needs.

\subsection{External Representations in Knowledge Work}
\label{sec:external-representations}

Knowledge work activities such as sensemaking and design require people to gather relevant materials, compare interpretations and alternatives, and revise intermediate judgments as their understanding develops~\cite{Russell1993_MLTWGDTN,Pirolli2005_X2GDEDPH,schon1983reflective,dorst2001creativity}. External representations participate in this work by shaping what information is perceptually available and what operations people can perform on it~\cite{norman1991cognitive,zhang1994representations,scaife1996external,hutchins1995cognition}. For example, representations of design rationale make issues, alternatives, and supporting arguments explicit, so that people can examine their relationships and reconsider earlier judgments~\cite{maclean1991questions,conklin1988gibis}.

Manipulating a representation can contribute to the work of understanding. Kirsh et al. describe epistemic actions as changes to the environment that simplify cognitive work~\cite{kirsh1994distinguishing}. In sensemaking, Wright et al. illustrate how arranging, grouping, and connecting materials can support the construction and revision of an analysis~\cite{Wright2006_672H2RF4}. Representational operations can also affect computation through different mechanisms. In semantic interaction, systems interpret users' manipulations as signals for updating an underlying model~\cite{endert2012semantic}, whereas Mackay et al. describe how systems maintain constraints and dependencies among organized objects during interaction~\cite{Mackay2025_InteractionSubstrates}.

Persistent representations make existing organization available to be revisited and revised. Studies of physical knowledge work show how location and visibility can mark relevance, priority, and unfinished activity~\cite{malone1983desks,kidd1994marks}. Research on spatial hypertext shows how organization can emerge through proximity, visual attributes, and lightweight relations~\cite{marshall1994viki,marshall1995spatial}. Such organization can remain available without becoming fixed. Beaudouin-Lafon et al. likewise emphasize that objects remain malleable over time as they are appropriated and reshaped in use~\cite{BeaudouinLafon2021_GenerativeTheories}. These accounts provide a foundation for examining how evolving representations organize reasoning and action in human--AI work.

\subsection{Interaction Models for Human--AI Work}
\label{sec:representational-workspaces}

Prior human--AI systems instantiate representation-centered interaction in different ways: they vary in what is represented, how that representation can be reorganized, and how its organization informs subsequent model action~\cite{Suh2023_5M75CFML,Kang2023_Synergi,Chen2025_MeetMap,Chen2026_ThinkFlow,Zhou2026_JCPKGCB6,Zhang2023_A9FG546C,Tang2026_LECYXG6A}. This variation motivates a broader account of how representation organizes continued work.

Several models explain how representations shape generative action. Shen et al. model the information and control flows through which prompts, interface actions, and artifacts combine into instructions for GenAI~\cite{Shen2026_IAI}. AI-Instruments embodies intent in manipulable, reusable instruments through which users steer generation and refine their intent~\cite{Riche2025_AIInstruments}. Kim et al. propose an object-oriented design framework that makes inputs, model configurations, and output spaces persistent, composable objects for assembling and comparing alternative generation configurations~\cite{kim_cells_2023}.

Other frameworks foreground coordination and adaptation in human--AI collaboration. The Co-Creative Framework for Interaction design characterizes turn-taking, initiative, communication, and contributions to a shared creative product~\cite{Rezwana2023_COFI}. Wang et al. propose a framework that makes the collaborative process explicit, persistent, and adaptable. It organizes problems, workflows, operations, environment, and reflection to support reframing goals, reorganizing dependencies, and redirecting ongoing work~\cite{Wang2025_IPI}.

Recent work focuses on persistent analytical artifacts and shared sensemaking representations. Figures as Interfaces links scientific figures to data, code, and provenance. Bidirectional mappings let visual interactions guide further analytical operations, while versioned artifacts preserve linked figures and their analytical provenance~\cite{Wang2026_FiguresAsInterfaces}. Guan et al. propose a dynamic shared workspace where humans and specialized agents organize evidence and interpretations across representational layers. Contributions are treated as candidates for negotiated incorporation, with competing interpretations and provenance kept visible~\cite{Guan2026_SharedRepresentationalWorkspace}.

Building on these accounts, we introduce the \textbf{Malleable Workspace}, a conceptual model of how an evolving Work Representation mediates human--agent collaboration. The model articulates how humans and agents form and revise represented organization, how that organization grounds subsequent action, and how such action changes the broader Work Environment. By making these relationships explicit, the model supports analysis of how this mediation operates and identifies opportunities for further exploration in the design of human--agent workspaces.

\section{From Structural Mismatches to the Malleable Workspace}
\label{sec-design}
This section characterizes the structural mismatches motivating the Malleable Workspace and presents its core constructs and relationships. The model presented here incorporates refinements informed by the cross-system analysis and longitudinal field study reported in Sections~\ref{sec-method} and~\ref{sec-study}.

\subsection{Structural Mismatches in Conversation-Centered AI Interaction}
\label{sec:structural-mismatch}

\begin{figure*}[tbp]
  \centering
  \includegraphics[width=\textwidth]{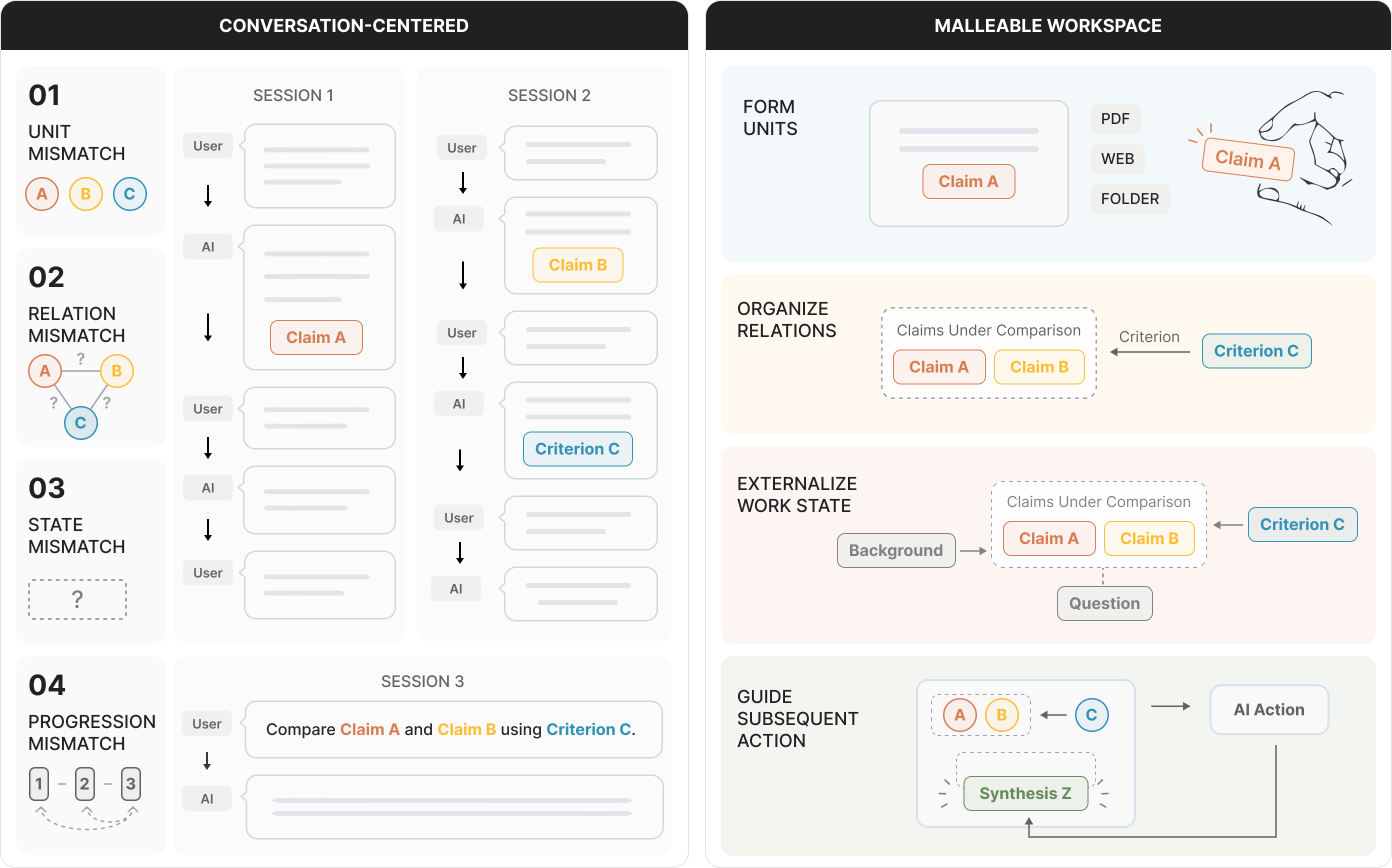}
  \caption{\textbf{Four structural mismatches in conversation-centered AI interaction and the corresponding capabilities of a Malleable Workspace.} Left: Claims A and B and Criterion C remain distributed across conversational turns and sessions, illustrating unit, relation, state, and progression mismatches; the user reconstructs the relevant context in a subsequent prompt. Right: A Malleable Workspace supports forming units from heterogeneous materials, organizing relations, externalizing the current work state, and using this representation to guide subsequent AI action. Action outcomes, such as Synthesis Z, can be incorporated into the evolving representation to support continued work.}
  \Description{A two-panel comparison of conversational histories and an evolving work representation. On the left, Claim A appears in Session 1, while Claim B and Criterion C appear in Session 2. In Session 3, the user brings these materials together in a comparison request. On the right, Claims A and B are grouped as units, with Criterion C linked to the group. An arrow leads from the representation to an artificial intelligence action, and the resulting Synthesis Z returns to the representation. The contrast is between reconstructing context across conversations and maintaining an organization that can guide further action.}
  \label{fig:structural-mismatch}
\end{figure*}

Consider a researcher studying why people struggle to keep track of their work as they explore ideas with AI. Across sessions and branches, she develops two candidate explanations and a criterion for comparing them, while relevant evidence remains scattered across the conversation history. To ask AI to compare them, she must include the explanations, the criterion, and the relevant evidence in a prompt. The comparison informs what evidence she seeks next and how she revises the manuscript. We characterize this structural mismatch between conversational progression and evolving work organization through four interrelated dimensions: \textbf{unit, relation, state, and progression} (Figure~\ref{fig:structural-mismatch}).

\textbf{Unit mismatch.} People may work with individual claims within responses or explanations assembled across turns and sources. As understanding develops, people may split an explanation or combine several claims. When prompts and responses are the primary addressable units, users must extract and recombine fragments to work with objects at other granularities.

\textbf{Relation mismatch.} Knowledge work involves forming and revising relations through which materials become meaningful---for example, whether one claim supports, contradicts, or qualifies another~\cite{suthers2003guidance,Wright2006_672H2RF4,Andrews2010_KH2GRBIV}. Conversation preserves chronological and branch structure, but not necessarily these evolving semantic, evidential, or developmental relations~\cite{herring1999coherence}.

\textbf{State mismatch.} Continued work depends on understanding which interpretations remain current, what is unresolved, and how tasks depend on one another~\cite{Russell1993_MLTWGDTN,Pirolli2005_X2GDEDPH}. Preserving earlier exchanges does not necessarily make this current work state explicit~\cite{bellotti2005quality,zhang2018groupsensemaking,sabei2024professionalim,smith2000conversationtrees,zhang2025mindalogue}.

\textbf{Progression mismatch.} Continued work depends on connecting materials, interpretations, and outcomes to subsequent action~\cite{bellotti2005quality,schmidt1996coordination}. When these span conversations and sources, continuing one conversational path does not necessarily make them available for the next action; users must assemble the relevant context to carry the work forward.

\subsection{The Malleable Workspace Model}
\label{sec:malleable-workspace}

Motivated by these mismatches, the \textbf{Malleable Workspace} centers on an evolving \textbf{Work Representation} through which humans and agents organize relevant elements and relations, revise that organization as work develops, and ground subsequent action within a broader \textbf{Work Environment} (Figure~\ref{fig:malleable-workspace-model}).

Within this representation, \textbf{units} are elements treated as coherent objects of work, whose boundaries need not coincide with source messages or documents; \textbf{relations} express how units are currently understood with respect to one another. The researcher, for example, can assemble two explanations across conversations and connect evidence to the explanations it supports or challenges. The represented state is expressed through this current organization, not a separate summary or fixed record. \textbf{Representational malleability} is the capacity of humans and agents to reshape the Work Representation as a way of advancing their work.

The \textbf{Work Environment} comprises the tools, source materials, artifacts, agents, and other resources involved in the work. The Work Representation draws on this environment to selectively externalize and organize what currently matters rather than reproducing it in full. For the researcher, this environment includes the source materials she consults, the manuscript she revises, and the tools and agents she uses.

\textbf{Human guidance} provides direction for a particular agent action, indicating how the Work Representation should be used without requiring users to reconstruct that context in each prompt.

The model specifies three relationships that can operate in different orders.

\textbf{Representational grounding (R1).} The Work Representation grounds human or agent action when its organization of units and relations is used as a basis for understanding the work and guiding action. In the researcher scenario, the agent draws on the represented explanations and how evidence supports or challenges them, while human guidance directs the agent to compare the explanations using the represented criterion.

\textbf{Selective interpretation and externalization (R2).} Humans and agents determine which materials and outcomes are relevant, interpret their significance for the current work, and externalize those interpretations by adding or revising units and relations. For the researcher, this can mean interpreting how an agent-generated comparison qualifies an explanation and updating the Work Representation accordingly, rather than merely retaining the output. As relevance and meaning change, units and relations may be reorganized or removed.

\textbf{Action-driven environmental transformation (R3).} R3 occurs when a human or agent action grounded in the Work Representation produces consequential changes in the broader Work Environment. For the researcher, this means using the comparison to revise the actual manuscript, rather than only updating represented explanations. Relevant consequences, such as feedback on the revision, may subsequently be interpreted and incorporated through R2.

\section{The Malleable Workspace as an Analytic Lens}
\label{sec-method}
We use the Malleable Workspace as a sensitizing framework to examine how evolving Work Representations participate in continued human--agent action. We reconstructed how systems organized work as units and relations, how that organization grounded particular agent actions, how agent contributions entered or revised represented work, and where action produced consequential change. The comparison identifies recurring configurations across interface forms, distinguishing how systems connect represented work to subsequent action and the boundaries of their support.

\subsection{Comparative Approach}
\label{sec:comparative-approach}

To construct the comparative corpus, we conducted a bounded search of main-proceedings full papers at CHI, UIST, CSCW, IUI, and DIS from 2022 through August 2026, supplemented by backward and forward citation tracing. We chose 2022 as the starting point to focus on systems built around recent generative models. We checked annual proceedings listings and metadata to identify candidate papers. Citation tracing admitted papers from other venues under the same eligibility criteria. Two authors divided the screening workload and cross-checked each other's decisions. Title and abstract screening retained papers plausibly reporting implemented GenAI systems for open-ended, multi-step knowledge work. Full texts were assessed against two conjunctive criteria: each system (1) maintained an inspectable and revisable representation functionally distinct from both a chat transcript and a final artifact, and (2) used human-selected or human-revised content or organization from that representation in at least one subsequent agent action. Public evidence also had to be sufficient to reconstruct this workflow. The comparison reported here covers 61 implementation cases from 58 source papers. Distinct implementations reported in the same paper were treated as separate cases when they organized represented work differently.

Using a provisional version of the model as a sensitizing framework, we reconstructed how agents participated in represented work across systems and developed a shared codebook. We recorded what became represented as units and relations, which aspects of this organization became available to particular agent actions, how generated material entered or changed represented work, and whether represented work informed changes to another object in the Work Environment. For each case, we coded a feature as supported when the paper explicitly reported at least one implemented mechanism meeting its definition. Two coders independently annotated a ten-system pilot. Discrepancies and proposed codebook revisions were discussed with three additional authors to clarify category boundaries and challenge initial interpretations. Following the pilot, the two coders stabilized the codebook for the main coding pass, divided the remaining corpus, and cross-checked each other's annotations, referring ambiguous cases to the five-author team. Analysis nevertheless remained iterative: cross-case comparison and difficult cases prompted targeted re-examination of earlier annotations, while subsequent refinements to the model led us to return to relevant cases before finalizing the analysis. Through this process, we developed the recurring configurations reported below.
Figure~\ref{fig:cross-case-matrix} summarizes these cross-case patterns alongside the visual forms used to represent work.

\begin{figure*}[!tbp]
  \centering
  \input{Table/system-coding-matrix-rows.tex}
  \begingroup
  \definecolor{HeaderNavy}{HTML}{213C51}
  \definecolor{FOne}{HTML}{3291B6}
  \definecolor{FTwo}{HTML}{658C58}
  \definecolor{FThree}{HTML}{FFBC37}
  \definecolor{FFour}{HTML}{D97D55}

  \newcommand{\codecore}[1]{%
    \cellcolor{#1!27}\textcolor{black!78}{\raisebox{0.05ex}{\(\checkmark\)}}%
  }
  \newcommand{\codenone}[1]{%
    \cellcolor{#1!2}\textcolor{black!28}{\textendash}%
  }
  \newcommand{\Code}[2]{\csname code#2\endcsname{#1}}
  \newcommand{\SystemCite}[1]{%
    \nobreak\hspace{0.12em}{\fontsize{4.2}{4.5}\selectfont\cite{#1}}%
  }
  \newcommand{\topheadstrut}{\rule[-1.5mm]{0pt}{5.4mm}}
  \newcommand{\subheadstrut}{%
    \rule[-\dimexpr\dp\strutbox+.35mm\relax]{0pt}{\dimexpr\ht\strutbox+\dp\strutbox+.7mm\relax}%
  }
  % Paint one continuous background from the second row, then the spanning label.
  % Symmetric second-row padding leaves the spanning label's center unchanged.
  \newcommand{\spanhead}[2]{%
    \cellcolor{HeaderNavy}\color{white}\subheadstrut
    \makebox[0pt][c]{\raisebox{0pt}[0pt][0pt]{\textcolor{HeaderNavy}{%
      \rule[-\dimexpr\dp\strutbox+.35mm\relax]{\dimexpr#1+2\tabcolsep\relax}{\dimexpr5.4mm+\ht\strutbox+\dp\strutbox+.7mm\relax}%
    }}}%
    \makebox[0pt][c]{\raisebox{\dimexpr2.7mm+.5\ht\strutbox-.5\dp\strutbox-.5\height+.5\depth\relax}[0pt][0pt]{\textbf{#2}}}%
  }
  \newcommand{\grouphead}[3]{%
    \multicolumn{#1}{c|}{%
      \cellcolor{#2}\color{white}%
      \makebox[0pt][c]{\topheadstrut\textbf{#3}}%
    }%
  }
  \newcommand{\subhead}[2]{%
    \cellcolor{#1!88}\color{white}\subheadstrut\textbf{#2}%
  }
  \newcommand{\legendheading}[2]{%
    \textcolor{#1}{\bfseries #2}\par%
  }
  \newcommand{\legenditem}[2]{\textbf{#1}\enspace #2\par}

  \newcolumntype{S}{%
    >{\fontsize{5.3}{5.6}\selectfont}w{l}{1.9 cm}%
  }
  \newcolumntype{V}{%
    >{\fontsize{5.3}{5.6}\selectfont}w{l}{1.05cm}%
  }
  \newcolumntype{C}{>{\centering\arraybackslash}m{0.847cm}}
  \newsavebox{\matrixbox}

  \setlength{\tabcolsep}{1.1pt}
  \renewcommand{\arraystretch}{0.87}
  \arrayrulecolor{black!12}

  \fontsize{6.5}{6.9}\selectfont
  \sbox{\matrixbox}{%
      \begin{tabular}{|S|V|*{13}{C|}}
        \hline
        \multicolumn{1}{|c|}{%
          \cellcolor{HeaderNavy}\topheadstrut
        } &
        \grouphead{1}{HeaderNavy}{} &
        \grouphead{6}{FOne}{F1: Representation} &
        \grouphead{3}{FTwo}{F2: Grounding} &
        \grouphead{2}{FThree}{F3: Incorp.} &
        \grouphead{2}{FFour}{F4: Reach} \\
        \cline{3-15}
        \multicolumn{1}{|c|}{\spanhead{1.9cm}{Case}} &
        \multicolumn{1}{c|}{\spanhead{1.05cm}{Form}} &
        \subhead{FOne}{Cnt} &
        \subhead{FOne}{Com} &
        \subhead{FOne}{Op} &
        \subhead{FOne}{Int} &
        \subhead{FOne}{Cmp} &
        \subhead{FOne}{Dev} &
        \subhead{FTwo}{Str} &
        \subhead{FTwo}{Cpt} &
        \subhead{FTwo}{Agt} &
        \subhead{FThree}{Loc} &
        \subhead{FThree}{Org} &
        \subhead{FFour}{WR} &
        \subhead{FFour}{Env} \\
        \hline
        \SystemCodingMatrixRows
      \end{tabular}%
  }
  \resizebox{\textwidth}{!}{%
    \begin{minipage}{\wd\matrixbox}
      \centering
      \usebox{\matrixbox}

      \vspace{0.8mm}
      \begin{minipage}{\linewidth}
        \begin{minipage}[t]{0.238\linewidth}
          \RaggedRight\fontsize{6.1}{6.3}\selectfont
          \legendheading{FOne}{F1: Work Representation}
          \legenditem{Cnt}{Substantive work content.}
          \legenditem{Com}{Commitments guiding the work.}
          \legenditem{Op}{Operation units.}
          \legenditem{Int}{Interpretive continuity.}
          \legenditem{Cmp}{Compositional continuity.}
          \legenditem{Dev}{Developmental continuity.}
        \end{minipage}\hfill
        \begin{minipage}[t]{0.238\linewidth}
          \RaggedRight\fontsize{6.1}{6.3}\selectfont
          \legendheading{FTwo}{F2: Representational Grounding}
          \legenditem{Str}{System structural assembly.}
          \legenditem{Cpt}{System computed assembly.}
          \legenditem{Agt}{Agent-directed acquisition.}
        \end{minipage}\hfill
        \begin{minipage}[t]{0.238\linewidth}
          \RaggedRight\fontsize{6.1}{6.3}\selectfont
          \legendheading{FThree}{F3: Incorporation}
          \legenditem{Loc}{Local incorporation.}
          \legenditem{Org}{Reorganization of existing units.}
        \end{minipage}\hfill
        \begin{minipage}[t]{0.238\linewidth}
          \RaggedRight\fontsize{6.1}{6.3}\selectfont
          \legendheading{FFour}{F4: Action Reach}
          \legenditem{WR}{Changes within the Work Representation.}
          \legenditem{Env}{Change to Work Environment.}
        \end{minipage}
      \end{minipage}
    \end{minipage}%
  }
  \endgroup
  \caption{Cross-case coding of \SystemCodingCaseCount{} implementation cases from \SystemCodingPaperCount{} papers. A checkmark indicates reported support; a dash indicates that support was not established from the available evidence. The Form column records the core visual forms used to organize and present work: Node-link, visibly linked nodes; Spatial, position and proximity, grouping, containment, or attribute-coordinate layouts; Table, discrete row-column cells; Document, continuous prose; Notebook, segmented inputs paired with in-place results; Mixed, multiple core forms with no single form dominant; Other, remaining forms, including lists and outlines. CGL's three prototypes and ITD's two implementations are coded separately.}
  \Description{A cross-case coding matrix with implementation cases as rows and representational and action-related features as columns. Among the 61 implementation cases, reported support counts for Work Representation are: substantive work content, 61; commitments guiding the work, 51; operation units, 21; interpretive continuity, 58; compositional continuity, 58; and developmental continuity, 29. For Representational Grounding: system structural assembly, 50; system computed assembly, 39; and agent-directed acquisition, 4. For Incorporation: local incorporation, 59; and reorganization of existing units, 22. For Action Reach: changes within the Work Representation, 59; and changes beyond the Work Representation, 10.}
  \label{fig:cross-case-matrix}
\end{figure*}

\subsection{Forming Work Representations through Units and Relations}
\label{sec:forming-work-representations}

\textbf{F1. Across diverse interface forms, systems differed in what they made addressable as units and which forms of continuity they preserved through relations.} Unitization made different aspects of work available for direct intervention, while relations retained how those elements were understood, belonged together, or had developed. These representational roles can coexist within a system. The meanings and organization of units and relations were often specified by task-specific workflows, while some systems left aspects of these open to user configuration.

\subsubsection{Systems Varied in What They Made Addressable as Units}

\textbf{Making substantive work content directly addressable.} Unitization turned content embedded in linear model output or a composite artifact into objects that could be manipulated independently. \textbf{Graphologue}~\cite{Jiang2023_IGH9MP27} converted concepts in streaming responses into entity nodes for targeted exploration. \textbf{Interactive Reasoning}~\cite{Pang2026_VZ3SIUBQ} represented a verbose reasoning chain as editable topic nodes and subtrees. Extending this move to a developing artifact, \textbf{Script\&Shift}~\cite{Siddiqui2025_VK3VMJ4M} organized draft content into layers that writers could develop separately.

\textbf{Making commitments guiding the work explicit and revisable.} Unitization separated what the work should satisfy from the content being produced. \textbf{OnGoal}~\cite{coscia2025ongoal} consolidated goals expressed across multi-turn dialogue into a goal model that users could review and control over time. \textbf{IntentTagger}~\cite{Gmeiner2025_ABHVR7PW} represented intent as editable tags with deck- or slide-level scope.

\textbf{Making operations inspectable and editable.} Unitization made operations themselves---not only their outcomes---available for direct intervention. \textbf{Spellburst}~\cite{Angert2023_2U6NHQVM} represented creative-coding transformations as operator nodes between sketch versions, exposing operations otherwise buried in editing history. \textbf{ThinkFlow}~\cite{Chen2026_ThinkFlow} transformed multi-turn sensemaking into editable workflow phases and dependencies that could be adapted across uses.

\subsubsection{Systems Varied in the Connections They Preserved through Relations}

\textbf{Preserving interpretive continuity.} Relations made interpretations of represented materials inspectable through their organization. \textbf{Orca}~\cite{Jiang2026_IFSGCMMF} allowed users to spatially arrange and group webpages according to distinctions relevant to the work, keeping categories and priorities visible as browsing continued. \textbf{Orality}~\cite{Li2026_2KQISMFJ} organized spoken ideas through topic structures and explicit conflict links that users could revise as their understanding developed.

\textbf{Preserving compositional continuity.} Relations kept local units connected to the larger structures they formed. \textbf{Sensecape}~\cite{Suh2023_5M75CFML} linked topic canvases through an abstraction hierarchy, allowing users to explore a subtopic while retaining its place in the broader information space. \textbf{Brickify}~\cite{Shi2025_HJG8Z7RT} composed design tokens into a visual lexicon through spatial and scoped relations, preserving how individual elements belonged within an intended image.

\textbf{Preserving developmental continuity.} Relations preserved how units developed from one another, making the development of represented work inspectable. \textbf{IdeationWeb}~\cite{Shen2025_EFGK8YRB} linked derived ideas to their sources through derivation paths, preserving how human- and AI-generated ideas evolved. \textbf{Spatial Balancing}~\cite{Fu2026_BJ5DCQUT} retained revisions as parent--child relationships in a version tree, enabling users to compare alternatives and return to earlier states.

\subsection{Grounding Agent Action in Represented Work}
\label{sec:representational-grounding}

\textbf{F2. Across systems, system-assembled context was common; agent-directed acquisition of further material during action was less common.} Human guidance identified a referent, scope, or operation; system mechanisms assembled additional context; and agents could use intermediate results to decide what further material to inspect or retrieve.

Users designated content from the Work Representation for agent action in different ways. In \textbf{StoryNode}~\cite{Qin2025_DENJ4X2B}, users selected node content for inclusion in the prompt; in \textbf{CoNode}~\cite{Zhou2026_JCPKGCB6}, users connected input blocks to an AI node. \textbf{System assembly} went beyond reading or serializing this designated material: it added or derived further context through structural and computational mechanisms. \textbf{Structural assembly} applied predefined scope rules, as in \textbf{CoMAP}~\cite{Li2026_ZH4L7AZY}, \textbf{IntentTagger}~\cite{Gmeiner2025_ABHVR7PW}, \textbf{EchoMind}~\cite{Chen2025_UC5JLJIC}, and \textbf{CoExploreDS}~\cite{Chen2025_3E6AFBWN}. For local refinement, CoMAP added immediate neighbors to the selected unit; for graph-wide assistance, the agent instead received conversation history and a serialized full graph. \textbf{Computed assembly} selected or derived additional context through retrieval, ranking, filtering, clustering, or inference. \textbf{Semantic Prompting}~\cite{Tang2026_LECYXG6A} inferred revision intent from workspace changes and spatial metadata, while \textbf{CoQuest}~\cite{Liu2024_4YCK2XDD} and \textbf{DiscipLink}~\cite{Zheng2024_HC227UX8} retrieved source materials from represented anchors. \textbf{MindTrellis}~\cite{Li2026_YFH7U76P} retrieved source material using the query and existing graph context.

\textbf{Agent-directed acquisition} allowed an agent to assess intermediate results and decide whether and how to seek further information. In \textbf{CoQuest}~\cite{Liu2024_4YCK2XDD}, \textbf{Cocoa}~\cite{Feng2026_Cocoa}, and \textbf{Orca}~\cite{Jiang2026_IFSGCMMF}, this involved acquiring source materials from the Work Environment. In Orca, acquisition remained visible and interruptible, allowing the agent to extend its grounding while preserving human control.

\subsection{Incorporating Agent Outputs into the Work Representation}
\label{sec:incorporating-agent-contributions}

\textbf{F3. Local incorporation of agent outputs was more widely supported than reorganization of existing units.} This distinction concerns the scope of supported changes, from developing particular work elements to revising the organization that guides subsequent work.

\textbf{Local incorporation.} Incorporation added or revised particular units and the relations tying them to their inputs, sources, or other specified targets. In \textbf{CoNode}~\cite{Zhou2026_JCPKGCB6}, generated outputs became blocks linked to their source inputs. \textbf{Qlarify}~\cite{Fok2024_9G6DRD7Z} inserted generated elaborations at human-specified anchors, while \textbf{IdeationWeb}~\cite{Shen2025_EFGK8YRB} created structured idea nodes and recorded source--derived paths. \textbf{Patchview}~\cite{Chung2024_NGLXLY4A} and \textbf{StoryEnsemble}~\cite{Suh2025_R6H4YV3E} revised existing elements or propagated changes across predefined dependencies.

\textbf{Reorganization of existing units.} Outputs were incorporated by revising how existing units were organized together. \textbf{ThemeViz}~\cite{Kang2025_FK9XX3VC} reassigned existing code--text pairs across themes, \textbf{PaperBridge}~\cite{Zhang2025_ZP5KNI7N} regenerated cluster themes and paper assignments, and \textbf{OnGoal}~\cite{coscia2025ongoal} merged or replaced goals within its goal model. \textbf{MindTrellis}~\cite{Li2026_YFH7U76P} modified its concept graph and hierarchy, \textbf{Orality}~\cite{Li2026_2KQISMFJ} restructured topic canvases, and \textbf{ThinkFlow}~\cite{Chen2026_ThinkFlow} updated workflow phases and dependencies. These changes operated through unit roles and relation types provided by task-specific system design or prior user configuration.

\textbf{Authorization and materialization.} Alongside the scope of change, systems differed in who authorized incorporation and how it was carried out. Agent outputs could remain outside the Work Representation until directly incorporated by a human (\textbf{StoryNode}~\cite{Qin2025_DENJ4X2B}, \textbf{Memolet}~\cite{yen2024memolet}), appear as system-formed candidates requiring human approval (\textbf{CoExploreDS}~\cite{Chen2025_3E6AFBWN}, \textbf{Polymind}~\cite{Wan2025_23TJBNSL}, \textbf{IntentTagger}~\cite{Gmeiner2025_ABHVR7PW}, \textbf{Criticality}~\cite{Chang2026_72ZMZZUT}), or enter the representation through predefined system rules (\textbf{Patchview}~\cite{Chung2024_NGLXLY4A}, \textbf{CoNode}~\cite{Zhou2026_JCPKGCB6}, \textbf{MindTrellis}~\cite{Li2026_YFH7U76P}, \textbf{IdeationWeb}~\cite{Shen2025_EFGK8YRB}, \textbf{OnGoal}~\cite{coscia2025ongoal}).

\subsection{Changing the Work Representation and the Broader Work Environment}
\label{sec:effects-of-agent-action}

\textbf{F4. More systems supported agent-driven changes within the Work Representation than changes to distinct artifacts or environments beyond it.} This distinction concerns whether action revised the units and relations constituting represented work or used that organization to change another object in the Work Environment.

\textbf{Changes within the Work Representation.} The developing object was itself part of the Work Representation, so representational revision could carry out substantive task work. In \textbf{VideOrigami}~\cite{Cao2025_NGEF9N4X}, \textbf{Spellburst}~\cite{Angert2023_2U6NHQVM}, \textbf{AI Chains}~\cite{Wu2022_XBAHNC7P}, and \textbf{Sensecape}~\cite{Suh2023_5M75CFML}, agent action created or revised represented artifacts, workflows, or knowledge structures that remained available for inspection, editing, and subsequent action. Represented work could also be handed off for use elsewhere: \textbf{CoMAP}~\cite{Li2026_ZH4L7AZY}, \textbf{Orality}~\cite{Li2026_2KQISMFJ}, and \textbf{OntoScope}~\cite{Zhao2026_5LAH67XW} supported such handoff through exports of documents, memos, or serialized artifacts. These exports made represented work available for subsequent use, but handing off that work was distinct from using it to guide agent changes to another work object.

\textbf{Changes beyond the Work Representation.} By contrast, agent action used represented organization to create or revise a distinct artifact or environment. In \textbf{VISAR}~\cite{Zhang2023_A9FG546C}, revisions to the argumentative outline guided updates to a prototype draft; in \textbf{Semantic Prompting}~\cite{Tang2026_LECYXG6A}, changes to spatial organization and annotations guided revisions to an existing report; and in \textbf{IntentTagger}~\cite{Gmeiner2025_ABHVR7PW}, revised intent tags guided updates to individual slides or the deck. These mechanisms allowed users to continue shaping another work object by revising the Work Representation. \textbf{Orca}~\cite{Jiang2026_IFSGCMMF} used the organization of webpages on its canvas to act on live webpages embedded in the interface: sharing an interface did not make the webpages and the organization guiding action on them the same work object.

\section{MalleableSpace: From Instantiation to Model Refinement}
\label{sec-study}
We constructed MalleableSpace as a general-purpose instantiation of the emerging Malleable Workspace model and conducted a three-month formative field deployment with seven experienced agent users. Whereas Section~\ref{sec-method} examined representational configurations, this study investigated how participants formed, revised, and used represented organization to guide action in self-selected work, and how this activity connected to resources and action in the wider Work Environment. We analyze participants' practices and the design iterations they informed to examine the model's operational implications and clarify its account of the Work Environment.

\subsection{MalleableSpace: Instantiating the Model}
\label{sec:malleablespace-instantiation}

\begin{figure*}[t]
  \centering
  \includegraphics[width=\textwidth]{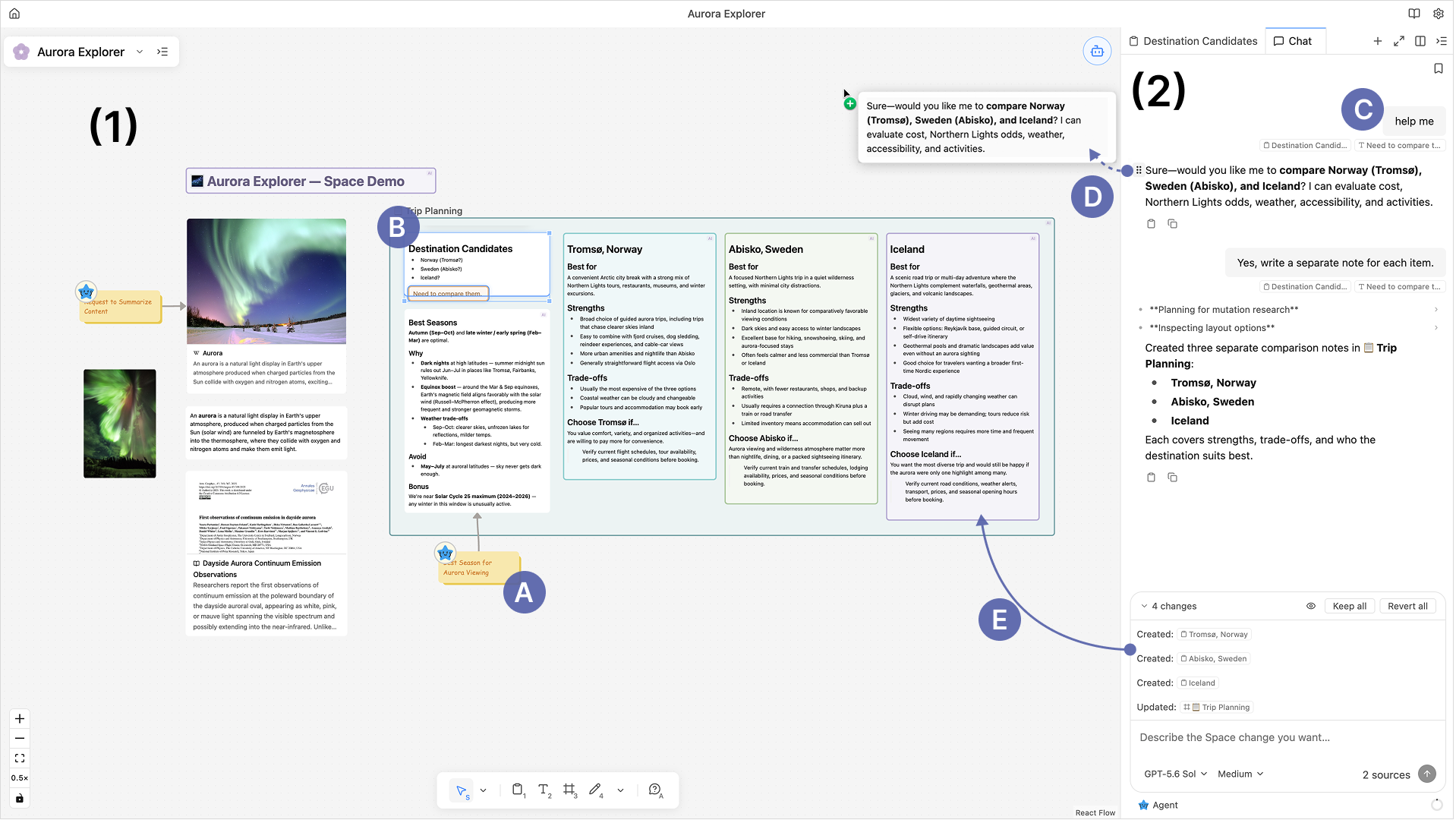}
  \caption{The MalleableSpace interface: (1) The Space Canvas holds documents, images, notes, and Agent Nodes; spatial arrangements, Frames, and labelled edges express relations among them. (2) The Side Panel supports agent conversations and viewing and editing individual units. (A) An Agent Node provides a persistent place on the canvas for an agent conversation, with connected units included as context. (B) Users select units to provide context and guidance for a particular request. (C) With this context available, a compact request can refer to represented work without restating it. (D) Users can drag a fragment of an agent response onto the canvas to create a unit for further organization and use. (E) In Agent mode, agents can edit unit content and reorganize units; users can review and keep or revert these changes.}
  \Description{A visual canvas on the left and a chat and content panel on the right show a destination-comparison task. Source images, documents, and framed notes surround Agent Nodes. Callout A marks connections between an Agent Node and other units. B highlights the selected destination-candidates note and comparison guidance; C shows the short request "help me" with attached context. D traces a response fragment being dragged from the side panel onto the canvas. E highlights generated notes and the associated change-review area.}
  \label{fig:malleablespace-interface}
\end{figure*}

MalleableSpace operationalized the Malleable Workspace model as an infinite visual Space with editable units and relations rather than a task-specific organizational scheme. Its \textbf{Space Canvas} hosted heterogeneous units---including Notes, Text nodes, PDFs, webpages, images, and videos---and let users revise their spatial arrangement, Frame membership, and labelled edges (Figure~\ref{fig:malleablespace-interface}(1)). Users could drag selected fragments from Chat responses, PDFs, and Notes onto the canvas (Figure~\ref{fig:malleablespace-interface}(D)), incorporating them into represented work \textbf{(R2)}. Together, these mechanisms supported the complementary operations examined in \textbf{F1}: making selected aspects of work addressable and organizing relations among them.

Users interacted with agents through general Chat or threads opened from persistent, spatially situated Agent Nodes in the Side Panel. User-selected units provided action-specific inputs and human guidance (Figure~\ref{fig:malleablespace-interface}(B)). For requests through an Agent Node, the system identified units near the node, connected to it by edges, or within the same Frame for the agent to inspect (Figure~\ref{fig:malleablespace-interface}(A)). Agents could inspect unit content and relations and search the Space for further materials, deciding what to read next as work on a request progressed. Grounding \textbf{(R1)} therefore extended user-selected inputs through structural assembly and agent-directed acquisition, bringing together mechanisms distinguished in \textbf{F2}. Users could refer to represented materials and organization in compact requests (Figure~\ref{fig:malleablespace-interface}(C)).

In \textbf{Ask mode}, agents returned responses in Chat, from which users could selectively externalize relevant fragments. In \textbf{Agent mode}, agents could revise unit content and associated relations, as well as reorganize existing units through changes to connections, Frame membership, and spatial arrangement. These operations supported both scopes of incorporation distinguished in \textbf{F3}: local changes and reorganization of existing units \textbf{(R2)}. Users could review agent changes and choose to keep or revert them (Figure~\ref{fig:malleablespace-interface}(E)). These initial capabilities focused on represented work within the Space; the later connection to user-owned agents is described in Sections~\ref{sec:field-deployment} and~\ref{sec:space-boundary}. MalleableSpace was implemented as an Electron desktop application with local, per-Space persistence and a built-in agent based on \texttt{pi-agent}. The source code for MalleableSpace is included in the supplementary material.

\subsection{Longitudinal Field Study}
\label{sec:field-deployment}

We conducted a three-month formative field deployment to understand how experienced agent users incorporated MalleableSpace into self-selected work, how their practices developed, and what breakdowns and design needs arose. The study was exploratory and qualitative, with participant feedback informing prototype development throughout the deployment. The protocol was approved by our institution's Institutional Review Board (IRB).

\subsubsection{Participants}

We recruited seven experienced agent users (P1--P7, four women and three men) with backgrounds in software development, product management, and HCI/AI research. All participants used LLM-based agents in their daily work and remained involved throughout the three-month deployment. Three had substantial prior experience with visual workspace tools such as Figma, while the others primarily used note-based tools such as Notion and Obsidian. We covered participants' monthly agent subscription costs.

\subsubsection{Study Procedure and System Iteration}

\textbf{Onboarding.} Participants completed a 30-minute onboarding session introducing system setup and core interactions, followed by a brief question-and-answer period. They also received a written guide for reference.

\textbf{Sustained use and reflection.} Participants used MalleableSpace in their own work on Windows or macOS computers, using their existing model subscriptions and API credentials; they had no assigned tasks or minimum-use requirements. During weekly one- to two-hour group reflection sessions, researchers presented recent system updates, followed by discussion of participants' uses, experiences, and desired improvements. The group format allowed participants to compare cases and surface emerging practices and design needs; a group chat supported questions and discussion between sessions.

\textbf{System iteration.} The deployment began with the built-in agent described in Section~\ref{sec:malleablespace-instantiation}. Participant feedback informed updates designed and implemented by the research team, most of which addressed bugs or local usability issues. To connect represented work with participants' existing tools and resources, we introduced support for user-owned agents in two stages. In Week 4, we added Agent Client Protocol (ACP) support so participants could connect their own agents; between Weeks 7 and 9, we progressively extended these agents' ability to inspect and modify the content and organization of the Space.

\textbf{Exit interviews.} At the end of the deployment, we conducted an approximately one-hour individual exit interview with each participant, asking them to reflect on the use cases they found most valuable, their overall experience with MalleableSpace, and opportunities for further improvement.

\subsubsection{Data and Analysis}

\begin{figure*}[t]
  \centering
  \includegraphics[width=\textwidth]{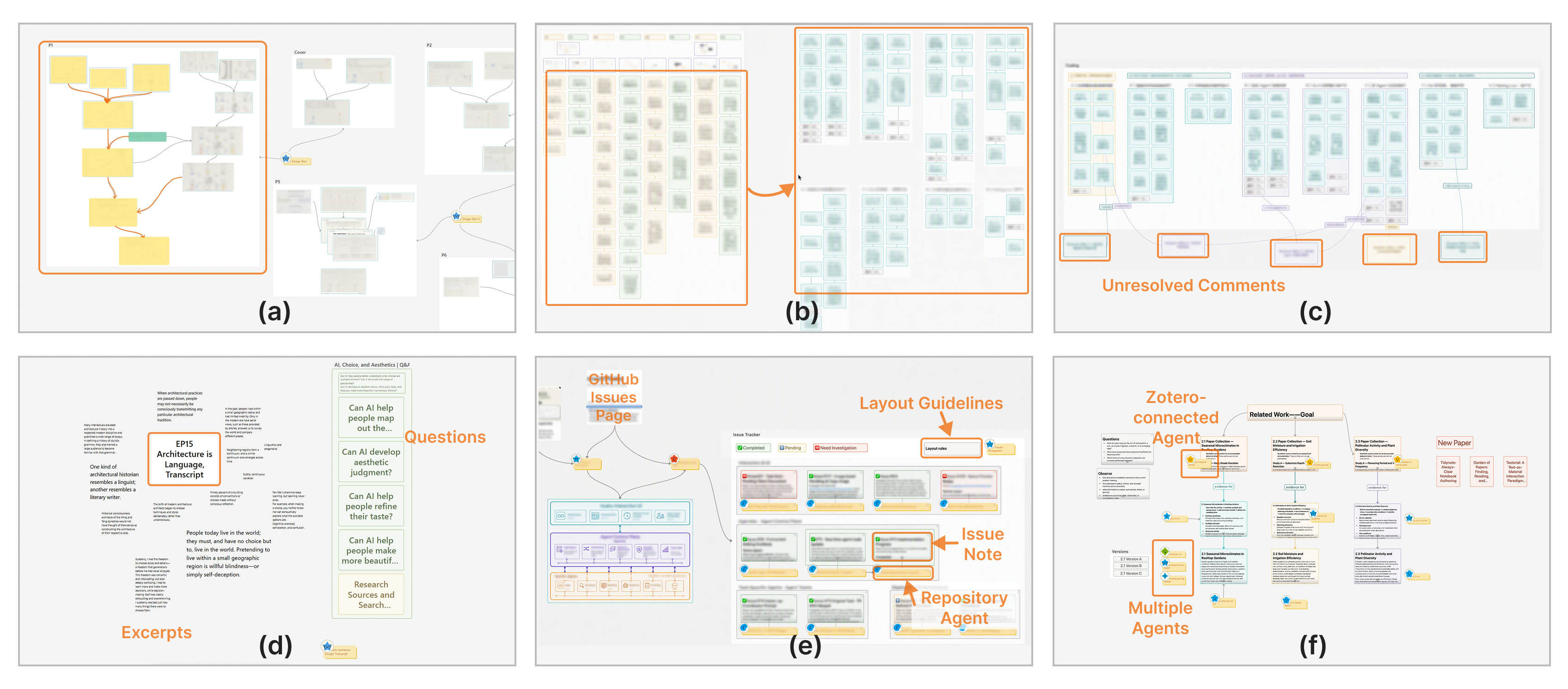}
  \caption{Snapshots from five participant-reported work episodes illustrating how represented organization developed and supported continued work. Orange outlines and arrows highlight key units and changes. (a) P2 developed several slide versions in parallel and guided an agent to combine selected elements before incorporating an additional source; yellow-highlighted image nodes and orange arrows trace this trajectory. (b--c) P5 collected interview findings in separate columns and asked an agent to organize them thematically (b). After finding the initial categories too loose, P5 and the agent revised them into three broader themes while leaving comments for further discussion visible (c). (d) P3 placed excerpts around a podcast transcript and organized the resulting questions in a column to the right for subsequent agent conversations. (e) P7 used a system architecture diagram and layout guidelines to organize software issues with a repository-grounded agent; colors marked status and edges expressed dependencies. (f) P1 used a Zotero-connected agent to retrieve papers from an existing collection and create three thematic notes in the Space, then explored alternative drafts with agents in different harnesses using the same represented content.}
  \Description{Six annotated panels, labelled a through f, are arranged in two rows. In a, arrows connect parallel slide versions to later combined versions. In b, an arrow leads from columns of interview material on the left to grouped findings on the right. In c, grouped findings appear above connected comments left for further discussion. In d, excerpts surround a central podcast transcript, with questions in a column to its right. In e, an architecture diagram and layout guidance accompany grouped issue nodes and an agent; node colors indicate status and edges indicate dependencies. In f, a goal at the top connects to three thematic literature branches, alongside draft versions and multiple agents. Some source text is blurred; the highlighted arrangements and connections remain visible.}
  \label{fig:participant-cases}
\end{figure*}

\textbf{Data corpus.} Our qualitative corpus comprised recordings and transcripts of the weekly group reflection sessions and seven exit interviews, the complete group chat history, and cases selected by participants from their Spaces. Transcripts were automatically generated and manually corrected. To protect participants' privacy and ownership of work-in-progress ideas, we did not access or analyze other Space content. Selected node content in the participant case snapshots (Figure~\ref{fig:participant-cases}) was blurred or reworded.

\textbf{Usage overview.} At exit interviews, participants reported having created a total of 43 Spaces during the deployment, ranging from 3 to 11 per participant. The largest reported Space contained 629 nodes. Participant-reported use cases included literature review, software development, presentation preparation, and personal knowledge work.

\textbf{Analysis.} We conducted an iterative, primarily inductive qualitative analysis alongside data collection. Two researchers reviewed the weekly materials, documented emerging observations as analytic memos, and discussed them with three additional researchers to challenge interpretations and refine the analytic focus. These observations served as sensitizing concepts for comparing subsequent cases rather than as fixed codes. After the deployment, the two researchers revisited the broader corpus, coded relevant excerpts, compared patterns across participants and sources, and developed themes, which were refined through full-team discussion and examination of disconfirming cases.

\subsection{Representational Roles and Evolving Organization}
\label{sec:roles-structure}

Findings from the deployment extend \textbf{F1} and \textbf{F3} by showing how participants gave units and relations work-specific roles and developed represented organization through human judgment and agent-assisted revision.

\subsubsection{Representational Roles Emerged in Use}

\textbf{Representational roles depended on how participants situated and used units and relations, rather than on node type alone.} P6 described using Text nodes to preserve unresolved decisions and prepare for further discussion:

\begin{quote}
``When I have explored two directions but have not yet decided which is more suitable, I record that thought and mark it with a salient color. I do not need to rewrite the question; I can select that node with the relevant materials and ask an agent to help me think through this decision.''
\end{quote}

The same Text node could serve two roles: a reminder of an unresolved decision and a prompt for discussing it with an agent. P3 similarly arranged podcast excerpts around a transcript, recorded questions prompted by them, and discussed those questions with an agent (Figure~\ref{fig:participant-cases}(d)). The transcript and excerpts provided source material and observations; their spatial arrangement expressed these relationships---a practice P3 said ``felt natural.''

For P7, represented material also specified how other represented work should be organized. A Frame of layout guidelines described how an agent should position and connect issue nodes during coding work (Figure~\ref{fig:participant-cases}(e)); colors marked issue status, edges indicated dependencies, and Frames grouped issues by architectural layer. The same Space thus held both issues and guidance for organizing them.

Across these cases, wording, visual treatment, placement, and relations distinguished materials to consult, questions to pursue, and guidance for action. These roles emerged through participants' use of the representation to preserve what mattered and direct subsequent work.

\subsubsection{Organization Evolved Through Human-Guided Agent Assistance}

\textbf{Organizational change involved participants assessing, redirecting, and sometimes deferring structure as their work developed.} P5's qualitative coding process illustrates how this judgment shaped agent-assisted revision (Figure~\ref{fig:participant-cases}(b--c)). P5 collected findings from individual interviews in separate columns, then asked an agent to organize them thematically. After judging the resulting categories too loose, P5 worked with the agent to consolidate those categories into three broader themes while retaining comments that marked issues for further discussion. The agent's initial grouping became a proposal to assess rather than a settled organization; P5's judgment of its adequacy directed the next revision.

P6 described abandoning an initial tree structure and leaving materials loosely arranged during exploration. Once a clearer organizing logic emerged, they articulated it to an agent to guide reorganization. P6 explained:

\begin{quote}
``Starting with structure is a shackle; it constrains my exploration. \ldots{} The Space became messier, but without those boundaries my thinking was free.''
\end{quote}

Structure also evolved through branching and recombination. While preparing slides, P2 developed several versions in parallel, then guided an agent to combine desirable elements from three versions and incorporate an additional source (Figure~\ref{fig:participant-cases}(a)). Earlier outputs became resources for a new composition, while alternatives remained available.

Across these cases, participants shaped the organization of units by assessing proposed structures, deciding when to leave materials loosely arranged, and selecting alternatives for recombination. Agents helped materialize these decisions through changes across existing units and relations.

\subsection{Connecting Represented Work and the Work Environment}
\label{sec:space-boundary}

\textbf{Continued use elaborated F4 by exposing a disconnection between organizing work in the Space and pursuing it with participants' existing resources and agents.} After two weeks, P1 reported having to re-upload papers and copy notes already maintained in Zotero. For P7, the difficulty operated in both directions: the repository was unavailable to the in-Space agent---``I cannot put the repository into this Space, so \ldots{} it does not understand my implementation''---and proposals developed there still had to be copied to a coding agent for implementation. Participants thus became context carriers, manually bringing materials into the Space and transferring represented conclusions elsewhere. Although the prototype supported the formation and revision of represented organization, using that organization to advance work on external resources and artifacts remained dependent on these manual transfers.

To address this disconnection, we progressively connected MalleableSpace to participants' user-owned agents through ACP, extending their ability to inspect and modify Space content and organization (Section~\ref{sec:field-deployment}). Agents could draw on resources in participants' existing Work Environments while working with represented materials and judgments in the Space. This enabled selective incorporation of external materials and allowed represented organization to guide requests for action on external artifacts, without requiring participants to reproduce their entire working environment in the Space.

\textbf{These connections allowed existing resources to enter represented work selectively and support its further development.} P1 connected a Zotero-enabled agent, which retrieved papers from an existing collection and organized them into three thematic notes in the Space (Figure~\ref{fig:participant-cases}(f)). P1 used these notes to develop a draft, then invited agents in different harnesses to explore alternatives from the same represented materials. P5 similarly asked Copilot and Codex to develop alternative plans for the same problem using a shared context. In these two cases, the agents' proposals were incorporated into the Space for comparison. External materials and agent contributions could thus become part of an organization that participants and agents continued to inspect and revise, rather than remaining separate from the work already developed there.

\textbf{Conversely, participants used represented organization to direct action on resources and artifacts beyond the Space.} P1 reported that, as they identified additional papers during revision, the Zotero-enabled agent added them to the external library, allowing work developed in the Space to inform changes to the collection on which later work would draw. P7's workflow made this direction especially visible. After linking GitHub issues into the Space, P7 organized and classified them in relation to the system architecture and discussed them with a repository-grounded agent. P7 reported that the agent then modified the local code according to the conclusions reached in the Space (Figure~\ref{fig:participant-cases}(e)). The relevant connection was not simply that an agent could access a repository: judgments developed through represented organization provided the basis for requests to change the implementation.

\textbf{The deployment informed both a system extension and a refinement of the model.} Connecting participants' own agents extended MalleableSpace's access to resources and action capabilities beyond the Space. In the reported cases, thematic notes and organized issues remained distinct from the Zotero collections and code repositories they drew on and helped participants change. These observations motivated the inclusion of the Work Environment as a core model component, encompassing both resources on which represented work depended and artifacts that could be changed through actions grounded in the Work Representation. The refinement therefore concerned how the Work Representation participated in ongoing work beyond its own revision, rather than simply expanding what the Space could contain.

\section{Discussion}
\label{sec-discussion}
\subsection{From Representing Work to Working Through Representation}
\label{sec:working-through-representation}

\subsubsection{Working On and Through Representation}

Research on external representations shows that selecting and reorganizing material can be part of reasoning and action, rather than merely documenting their outcomes~\cite{norman1991cognitive,zhang1994representations,kirsh1994distinguishing}. Human--AI research extends this account through representations that support instruction and steering~\cite{Shen2026_IAI,Riche2025_AIInstruments,Tang2026_LECYXG6A}, and the reuse of intermediate materials~\cite{Chen2026_ThinkFlow,Li2026_YFH7U76P}.

Our cross-system analysis and longitudinal field study suggest two connected ways of working with a Work Representation. Humans and agents work \textbf{on} it when they form and revise its organization: selecting what becomes a unit, establishing or revising relations, and reorganizing materials as their understanding changes. Here, the representation is an object of work, and changing its organization can itself be part of developing an understanding of the problem.

Humans and agents work \textbf{through} the Work Representation when its current organization guides their subsequent action. In our longitudinal field study, extending this way of working beyond the Space initially relied on participants carrying context between the Space and their existing tools and agents. Connecting the Space to these agents enabled participants to use represented organization directly when changing external artifacts, while relevant materials and outcomes could be selectively incorporated through further work \textbf{on} the representation. These observations motivated the inclusion of the Work Environment as a core model component and clarified R3: actions grounded in the Work Representation can modify that environment. Working on and through representation can thus form a recursive loop, but the transitions are not automatic: \textbf{visible does not imply grounded, generated does not imply incorporated, and represented does not imply consequential.}

\subsubsection{Malleability in Continued Human--Agent Work}

When represented organization guides action, changes in understanding may require revising not only individual units but also how those units are related and organized. Keeping such organization open to revision has a precedent in spatial hypertext's support for emergent structure~\cite{marshall1994viki,marshall1995spatial}. Our cross-system analysis found that more systems supported local incorporation of agent outputs than reorganization of existing represented work. Our longitudinal field study showed how participants recruited agents for such organizational transformations, regrouping, relating, and recombining existing units as the work developed. \textbf{Malleability thus concerns supporting changes at the scope required by the work, from local updates to coordinated revisions across units and relations.}

\textbf{Under human guidance, agents can help make representational malleability more practicable.} Our longitudinal field study highlights three connected forms of assistance. First, agents can interpret and synthesize heterogeneous materials into work-relevant candidate units. Second, agents can form and revise relations and reorganize units as understanding develops, rather than only generate content within a fixed structure. Third, agents can act on an explicit organizing scheme, translating its instructions into changes to the Work Representation. Such guidance may be supplied through reusable instructions (e.g., a skill) or notes within the Space. When these instructions are included in the context grounding an agent's action, they can guide subsequent organization; placing them as notes in the Space also keeps the guidance represented and revisable. These capabilities can thus support the continued formation and reorganization of the Work Representation, rather than only the production of individual outputs.

Seen in this way, a \textbf{Malleable Workspace} is a continuing medium in which humans and agents can revise represented work and use it to shape further activity. Although MalleableSpace explored this possibility through a spatial canvas, the model concerns the relationships among representation, action, and environment rather than a particular visual interface category.

\subsection{Design Considerations for Malleable Workspaces}
\label{sec:design-considerations}

Drawing on the Malleable Workspace model (Section~\ref{sec-design}) and recurring patterns from the cross-system analysis (Section~\ref{sec-method}) and longitudinal field study (Section~\ref{sec-study}), we identify four design considerations for Malleable Workspaces.

\subsubsection{DC1. How Should Work Be Made Addressable through Units and Relations?}

Designers should decide what aspects of ongoing work become independently addressable as units and which relations preserve how those units are understood and composed. Existing interaction containers---such as conversations, responses, or documents---do not always provide suitable boundaries: relevant content, commitments, and intermediate states may need to be separated from larger sources or assembled across them~\cite{Suh2023_5M75CFML}. Unit boundaries and relations may be predefined by domain schemas, formed directly by humans, or proposed by agents in response to higher-level intentions. In particular, agents can serve as an interpretive layer across heterogeneous sources, turning varied materials into candidate units and relations without requiring a dedicated integration for every source. Designers must therefore consider how units can be extracted at different granularities, represented and modified, while keeping agent interpretations inspectable. Stable schemas provide clarity, while open organization supports emerging understanding~\cite{marshall1994viki}.

\subsubsection{DC2. How Should a Work Representation Ground Particular Agent Actions?}

Designers should specify how a Work Representation grounds an agent action and how human guidance shapes that grounding. A human may identify a referent, scope, operation, or constraint through selections and other interface actions~\cite{Shen2026_IAI,Riche2025_AIInstruments,Tang2026_LECYXG6A}. Guidance may also persist within the representation: explicit instructions placed alongside work materials can specify information needs and organizational requirements across subsequent actions. Forming action-specific context may combine human selection, predefined rules for including related units, relations, attributes, and source materials, and agent-directed information seeking. Under human-specified goals, an agent may actively seek relevant information from both the Work Representation and the broader Work Environment rather than relying solely on context supplied in advance~\cite{Jiang2026_IFSGCMMF}. Designers should therefore decide what context to specify in advance, how far an agent may seek information beyond that context, and how humans can inspect and adjust both the scope of information seeking and the context used for an action. Broader context can preserve relevant connections but introduce irrelevant material; narrower context improves predictability but may omit information needed to interpret or act on selected units.

\subsubsection{DC3. How Should Agent Outputs Be Incorporated into the Work Representation?}

Designers should determine how agent outputs are taken up in continued work. An output may remain transient or become part of the Work Representation, including as a candidate for further consideration; incorporation means updating the represented work, not merely preserving a response. It may create or revise units, make relations explicit, or reorganize existing materials. In our cross-system analysis, local incorporation was widely supported, whereas fewer systems supported reorganization of existing units. Agents might propose coordinated changes or alternative organizations, while humans inspect and authorize them at different granularities. Designers should therefore consider how outputs are assessed for incorporation, how the scope of changes to the Work Representation is determined, and who approves and applies those changes. Local updates offer predictability, whereas broader reorganization may better reflect a developing understanding but risks disrupting established structure.

\subsubsection{DC4. How Should a Work Representation Stay Connected to the Work Environment?}

Designers should not treat a Work Representation as a self-contained space detached from the environment in which work unfolds. Although necessarily selective, it should remain connected to the external resources, capabilities, and objects of work that give represented materials meaning and consequence. Relevant external materials and changes may be incorporated to update the represented work~\cite{Gmeiner2025_ABHVR7PW,Jiang2026_IFSGCMMF}; conversely, units, relations, and their organization may ground actions on external artifacts~\cite{Tang2026_LECYXG6A,Gmeiner2025_ABHVR7PW,Jiang2026_IFSGCMMF}. The outcomes of those actions can return to the representation for further interpretation. Our deployment showed that without connections in both directions, humans became context carriers who manually transferred information into and out of the Space. Designers should therefore decide what may cross the representational boundary, how these transitions are authorized and inspected, and how represented state remains current without indiscriminate synchronization.

\subsection{Limitations and Future Work}
\label{sec:limitations-future-work}

First, applying the Malleable Workspace model to 58 prior papers provides evidence of its descriptive and analytical coverage across this corpus. Its generative and practical value as a design framework, however, remains to be further examined. Future work can investigate whether independent designers can apply the model consistently to identify meaningful design opportunities and translate them into concrete system designs. Design studies and cross-system implementations can assess how the model supports the generation, comparison, and refinement of interaction designs, as well as whether model-informed designs improve coordination, controllability, and task outcomes. Such investigations can further clarify the model's constructs and boundary conditions and inform extensions for multi-user, multi-agent, or temporally distributed work.

Second, our three-month deployment with seven experienced agent users provides initial empirical evidence of how the model can be instantiated and appropriated in sustained work. It revealed how people developed representational practices and encountered breakdowns when using MalleableSpace, but the participant group and reported use cases covered a relatively narrow range of users and work contexts. Future work can broaden this empirical basis by involving larger and more diverse populations and deploying MalleableSpace across additional occupations, tasks, and real-world settings. These studies could identify both recurring and new usage patterns, examine how experiences vary across users and forms of work, and provide stronger evidence for the model's practical value and boundary conditions.

\section{Conclusion}
\label{sec-conclusion}
Our findings suggest that AI-supported knowledge work can proceed not only through exchanges but also through an evolving representation that organizes what matters and guides what happens next. The Malleable Workspace conceptualizes this representation as a revisable medium connecting human and agent action to a broader Work Environment. The cross-system analysis demonstrated the model's comparative and diagnostic value, while the MalleableSpace study elaborated and refined the model through sustained use. Together, these investigations suggest that systems should sustain a revisable Work Representation through which evolving work organization can persist, ground subsequent action, and connect to the broader Work Environment. Beyond any single interface form, the Malleable Workspace points toward a representation-centered interaction paradigm in which persistent, revisable representations help people maintain and reshape work while grounding agent assistance in how that work is organized.

% \begin{acks}
% \end{acks}

% \section*{Ethics and Privacy Statement}

% This section of your ACM work should discuss the potential societal
% risks that might result from its publication; two to three sentences
% related to the findings of your study, or new advancements made
% possible by their developed methods. The privacy and ethics statement
% should clearly address the broader impacts of their work as it relates
% to the authors' interpretation of privacy, fairness, safety, human
% rights, data sovereignty, or future misuse and any benefit/risk
% trade-off resulting from this research. We acknowledge that some
% papers may have minimal societal risks beyond those considered by
% institutional review boards, and the dimensions considered by any
% review of the user study design or dataset licenses could be provided
% in this statement.

\bibliographystyle{ACM-Reference-Format}
\bibliography{main}

% \appendix









\end{document}